\documentclass[11pt]{article}

\usepackage[top=50mm, bottom=50mm, left=50mm, right=50mm]{geometry}
\usepackage{lineno}
\usepackage{amssymb}
\usepackage{amsmath}
\usepackage{amsthm}
\usepackage{epsfig}
\usepackage{graphicx}
\usepackage{graphics}
\usepackage{float}
\usepackage{subfigure}
\usepackage{multirow}
\usepackage{color}
\usepackage{lineno}
\usepackage{fullpage}
\usepackage[normalem]{ulem} 
\usepackage{makeidx}
\usepackage{xspace}
\usepackage{wrapfig}
\usepackage{todonotes}
\usepackage[linesnumbered,lined,boxed,commentsnumbered]{algorithm2e}
\usepackage{booktabs}
\usepackage{graphicx}
\usepackage{hyperref}
\usepackage[utf8]{inputenc}
\usepackage{mathtools}
\usepackage{multirow}
\usepackage{todonotes}
\usepackage{comment}
\usepackage{url}

\makeindex

\definecolor{darkred}{rgb}{1, 0.1, 0.3}
\definecolor{darkblue}{rgb}{0.1, 0.1, 1}
\definecolor{darkgreen}{rgb}{0,0.6,0.5}

\newcommand {\mm}[1] {\ifmmode{#1}\else{\mbox{\(#1\)}}\fi}

\begin{document}

\title{Minimod: A Finite Difference solver for Seismic Modeling}
 
\author{
{Jie Meng\thanks{Total EP R\&T, email: jie.meng@total.com}} \and Andreas Atle\footnotemark[1] \and {Henri Calandra\thanks{Total S.A.}} \and {Mauricio Araya-Polo\footnotemark[1]}
}

\maketitle
\setcounter{page}{1}

\begin{abstract}
This article introduces a benchmark application for seismic modeling using finite difference method, which is named \emph{MiniMod}, a \emph{mini} application for seismic \emph{mod}eling. The purpose is to provide a benchmark suite that is, on one hand easy to build and adapt to the state of the art in programming models and changing high performance hardware landscape. On the other hand, the intention is to have a proxy application to actual production geophysical exploration workloads for Oil \& Gas exploration, and other geosciences applications based on the wave equation. From top to bottom, we describe the design concepts, algorithms, code structure of the application, and present the benchmark results on different current computer architectures. 
\end{abstract}

\section{Introduction}
\label{sec:intro}
\emph{Minimod} is a Finite Difference-based proxy application which implements seismic modeling (see Chapter~\ref{sec:seimod}) with different approximations of the wave equation (see Chapter~\ref{sec:fd}). Minimod is self--contained and designed to be portable across multiple High Performance Computing (HPC) platforms. The application suite provides both non--optimized and optimized versions of computational kernels for targeted platforms (see Chapter~\ref{sec:description}). The target specific kernels are provided in order to conduct benchmarking and comparisons for emerging new hardware and programming technologies.\\

\noindent Minimod is designed to:
\begin{itemize}
\item Be portable across multiple software stacks. 
\item Be self--contained.
\item Provide non--optimized version of the computational kernels.
\item Provide optimized version of computational kernels for targeted platforms.
\item Evaluate node--level parallel performance.
\item Evaluate distributed--level parallel performance.
\end{itemize}
\noindent The first four items are covered in Section~\ref{sec:description} and the remainder items are covered in Section~\ref{sec:benchmarks}.\\

\begin{figure}[ht!]
    \centering
    \includegraphics[scale=0.5]{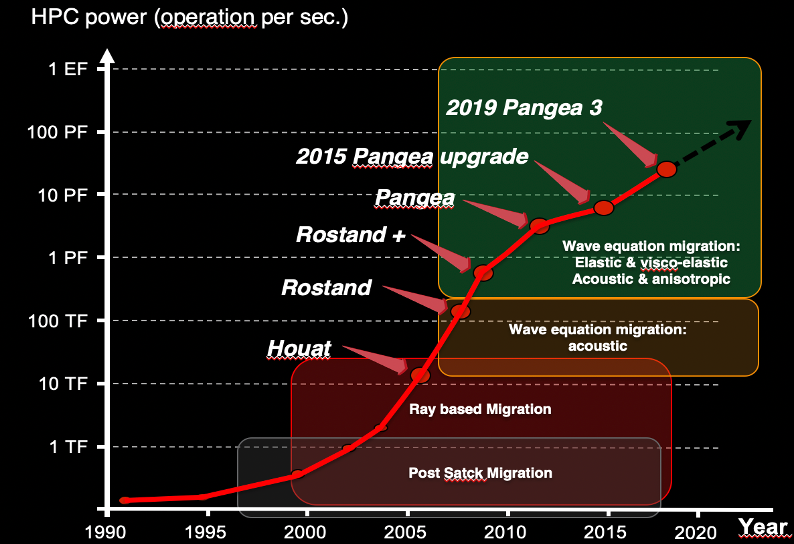}
    \caption{Evolution of computing capacity wrt geophysical algorithms.}
    \label{fig:evol}
\end{figure}

New HPC technologies evaluation is a constant task that plays a key role when decisions are taken in terms of computing capacity acquisitions. Evaluations in the form of benchmarking provide information to compare competing technologies wrt relevant workloads. Minimig is also use for this purpose, and insight collected with it has been part the last major acquisitions by Total (see Figure~\ref{fig:evol}). 

It can be observed in Figure~\ref{fig:evol}, how more complex geophysical algorithms drive larger capacity installations in Total. The main driver of performance demand by the geophysical algorithms presented in figure are: the accuracy of the wave equation approximation and the addition of optimization or inverse problem schemes. The former is captured in Minimig, where the later is out of scope of this article.
Performance trends obtained by conducting experiments with Minimig (or similar tools) influenced the decisions for the last ten years, this mainly motivated by the transition of the main workloads from Ray-based to wave-based methods.

\newpage

\section{Seismic Modeling}
\label{sec:seimod}
\begin{figure}[ht!]
    \centering
    \includegraphics[scale=0.5]{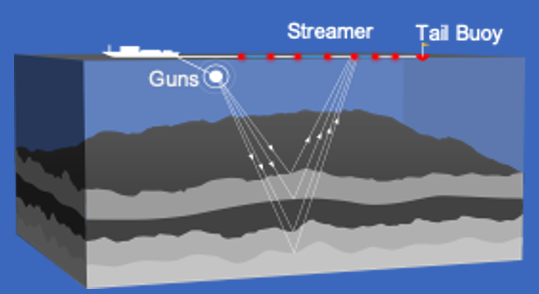}
    \caption{The mechanical medium represented by the subsurface model is perturbed and the wave propagation simulated. In this case, waves are depicted in discretized form as rays, for simplicity.}
    \label{fig:modeling}
\end{figure}

Seismic depth imaging is the main tool used to extract information describing the geological structures of the subsurface from recorded seismic data, it is effective till certain depth after which it becomes inaccurate. At its core it is an inverse problem which consists in finding the best model minimizing the distance between the observed data (recorded seismic data) and the predicted data (produced by computational means). The process to estimate the predicted data is known as \emph{forward modeling}. It is based on the resolution of the wave equation for artificial perturbations of the subsurface given initial and boundary conditions. This simulation is repeated as many times as perturbations were introduced during seismic data acquisition. In Figure~\ref{fig:modeling} on of such experiments is represented, in this case for a marine setup. The perturbation (namely \emph{source}) is introduce by an airgun dragged behind a ship, then the waves propagate through the medium. At each interface between layers of materials with different characteristics part of the energy is reflected. These reflections are recorded at sea level (at surface for a onshore setup) by a network of sensors (in the figure depicted in red) also pulled by the ship.

Solving the forward modeling efficiently is crucial for geophysical imaging as one needs to get solutions for many sources and many iterations as we progressively the subsurface model improves. Constant progresses in data acquisition and in rocks physics labs, more powerful computers and integrated team including physicists, applied mathematicians and computer scientists have greatly contributed to the development of advanced numerical algorithms integrating more and more complex physics. For the last 20 years, the field has been very active in the definition and introduction of different  wave equation approximations  and corresponding numerical methods for solving forward problem. But the real change came  with the implementation of the full wave equation, thanks to the petascale  era in HPC, giving access to a complete representation of the wavefield. It allowed geo-scientist to re-design imaging algorithm both in time dynamic and time harmonic domain. The most popular numerical scheme used, nowadays by the industry, is based on finite difference  methods (FDM) on regular grids \cite{kelly1976}, \cite{dablain1986}. We refer to \cite{virieux2009} for examples of FDM in the geophysics frameworks and to \cite{graves1996} for 3D applications.

\newpage

\section{Finite Differences}
\label{sec:fd}

Various numerical methods have been explored for the modeling of seismic wave propagation including the finite difference, finite element, finite volume and hybrid methods. Among those methods, the finite difference method is the most popular one for its simplicity and easy and straightforward implementation.\\

The first step of implementing the governing equations is to called discretizations, basically consist on write the equations on forms that allow the direct implementation of differential operators.  
The discretizations of the governing equations are impose on three different kind of grids, depending on the symmetry of the problem. We use the standard collocated grid, and two versions of staggered grid, namely Yee \cite{yee1966}, \cite{virieux1984}, \cite{virieux1986} and Lebedev \cite{lebedev1964}. \\

The first equation to be described is the second order acoustic wave equation with constant density, solving for the pressure wavefield $p$,
\begin{equation}
\frac{1}{v_p^2}\frac{\partial^2 p(\mathbf{x},t)}{\partial t^2} - \nabla^2 p(\mathbf{x},t) = f(\mathbf{x},t),
\label{acoustic_iso_cd}
\end{equation}
where $v_p$ is the velocity of the pressure wavefield, $p(\mathbf{x},t)$ expanded to 3D domain is $p(x,y,z,t)$, likewise for the source $f(\mathbf{x},t) = f(x,y,z,t)$.\\

The second equation is the first order acoustic wave equation with variable density $\rho$,
\begin{equation}
\frac{1}{\rho v_p^2}\frac{\partial p(\mathbf{x},t)}{\partial t} - \nabla\cdot\mathbf{v}(\mathbf{x},t) = f(\mathbf{x},t), \quad \rho\frac{\partial \mathbf{v}(\mathbf{x},t)}{\partial t} - \nabla p(\mathbf{x},t) = 0,
\label{acoustic_iso_vd}
\end{equation}
where $p$ is the pressure wavefield, and $\mathbf{v}$ is a vector wavefield for the particle velocities (time derivatives of displacement) along the different coordinate axis. \\

The third equation is the acoustic transversely isotropic first order system, see \cite{bube2012} for details.\\

Finally, we have the elastic equations with variable density $\rho$,
\begin{equation}
\frac{\partial \boldsymbol{\sigma}(\mathbf{x},t)}{\partial t} - CD\mathbf{v}(\mathbf{x},t) = \mathbf{f}(\mathbf{x},t), \quad
\rho\frac{\partial \mathbf{v}(\mathbf{x},t)}{\partial t} - D^t\boldsymbol{\sigma}(\mathbf{x},t) = 0,
\label{elastic_vd}
\end{equation}
where $\boldsymbol{\sigma}$ is a vector wavefield for the stresses using Voigt notation and $\mathbf{v}$ is a vector wavefield for the particle velocities.
The derivative operator $D$ is
\begin{equation}
D = \left(\begin{array}{ccc}
\frac{\partial}{\partial x} \\
& \frac{\partial}{\partial y} \\
&& \frac{\partial}{\partial z} \\
& \frac{\partial}{\partial z} & \frac{\partial}{\partial y} \\
\frac{\partial}{\partial z} && \frac{\partial}{\partial x} \\
\frac{\partial}{\partial y} & \frac{\partial}{\partial x}
\end{array}\right),
\label{d_operator}
\end{equation}
and $D^t$ is the transpose of $D$ without transpose of the derivatives. This is a subtle difference since a derivative is anti-symmetric. We have two different symmetry classes, isotropic and transversely isotropic, which only differs in the sparsity pattern of the stiffness tensor $C$.\\

The above described discretizations are implemented with the following names as kernels:
\begin{itemize}
\item Acoustic\_iso\_cd: Standard second order acoustic wave-propagation in isotropic media with constant density.
\item Acoustic\_iso: first order acoustic wave-propagation in isotropic media on a staggered Yee-grid variable density.
\item Acoustic\_tti: first order acoustic wave-propagation in transversely isotropic media on a staggered Lebedev-grid.
\item Elastic\_iso: first order elastic wave-propagation in isotropic media on a staggered Yee-grid.
\item Elastic\_tti: first order elastic wave-propagation in transversely isotropic media on a staggered Lebedev-grid.
\item Elastic\_tti\_approx: Non-standard first order elastic wave-propagation in transversely isotropic media on a staggered Yee-grid
\end{itemize}
All discretizations use CPML \cite{komatitsch2007} at the boundary of the computational domain, with the option of using free surface boundary conditions at the surface.
Full unroll of the discretization is given for \emph{acoustic\_iso\_cd}, as example, this is the simplest kernel in Minimod for simulating acoustic wave-propagation in isotropic media with a constant density domain , i.e. equation (\ref{acoustic_iso_cd}). The equation is discretized in time using a second-order centered stencil, resulting in the semi-discretized equation:

\begin{equation}
p^{n+1} - Qp^n + p^{n-1} = \left(\Delta t^2\right) v_p^2 f^n,
\label{eq:minimod-semidisc}
\end{equation}
where
\begin{equation*}
	Q = 2 + \Delta t^2 v_p^2 \nabla^2.
\end{equation*}

The equation is discretized in space using a 25-point stencil in space, with nine points in each direction of three dimensions:
\begin{align*}
\nabla^2 p(x,y,z) \approx \sum_{m=1}^4 &c_{xm}\left[p(i+m,j,k) + p(i-m,j,k) - 2p(i,j,k)\right] &+ \\
									 &c_{ym}\left[p(i,j+m,k) + p(i,j-m,k) - 2p(i,j,k)\right] &+ \\
									 &c_{zm}\left[p(i,j,k+m) + p(i,j,k-m) - 2p(i,j,k)\right]
\end{align*}
where $c_{xm}, c_{ym}$ and $c_{zm}$ are discretization parameters that approximates second derivatives in the different spatial directions.
The parameters can be derived from the Taylor expansion of the derivatives in the x, y and z-direction respectively, where the approximation would be of order 8. The derivatives can also use optimized stencils, that reduce the dispersion error at the expense of formal order.

\newpage

\section{Computing costs}
\label{sec:compcost}
Being the core algorithm of Finite Difference, stencil-based computation algorithms represent the kernels of many well--known scientific applications, such as geophysics and weather forecasting. \\

\begin{figure}[ht!]
    \centering
    \includegraphics[scale=0.25]{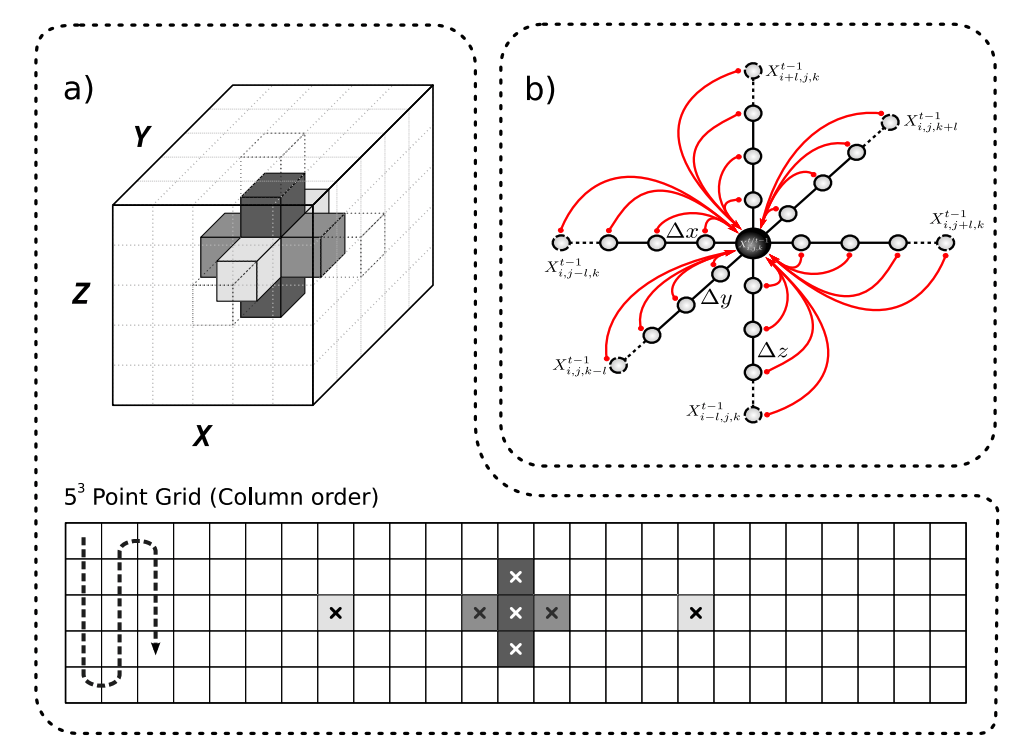}
    \caption{Memory layout for an simple stencil example, to access the required values multiple useless cache lines (bottom) need to be accessed with incurred penalties. Figure extracted from \cite{delacruz2014semi}.}
    \label{fig:memlay}
\end{figure}

However, the peak performance of stencil-based algorithms are limited because of the imbalance between computing capacity of processors units and data transfer throughput of memory architectures. In Figure~\ref{fig:memlay} the memory access problem is shown.
The computing part of the problem is basically the low re-use of the memory accessed elements.

In order to deal with the above described limitation, a great amount of research have been devoted to optimize stencil computations to achieve higher performance. For example, de la Cruz and Araya--Polo proposed the semi--stencil algorithm \cite{delacruz2014semi} which improves memory access pattern and efficiently reuses accessed data by dividing the computation into several updates. Rivera et al.~\cite{Rivera2000} showed that tiling 3D stencil combined with
array padding could significantly reduce miss rates and achieve performance improvements for key scientific kernels. Recently, Nguyen et al.~\cite{Nguyen2010} introduced higher dimension cache optimizations. \\

Advanced programming models have been explored to improve stencil performance and productivity. In 2012, Ghosh et al. \cite{ghosh2012experiences} analyzed the performance and programmability of three high-level directive-based GPU programming models (PGI, CAPS, and OpenACC) on an NVIDIA GPU for kernels of the same type as described in previous sections and for Reverse Time Migration (RTM, \cite{Araya2011}), widely used method in geophysics. In 2017, Qawasmeh et al. \cite{qawasmeh2017performance} implemented an MPI plus OpenACC approach for seismic modeling and RTM. Domain--specific languages (DSLs) for stencil algorithms have also been proposed. For example, Louboutin et al. introduced Devito~\cite{devito-api},  which a new domain-specific language for implementing differential equation solvers. Also, de la Cruz and Araya--Polo proposed an accurate performance model for a wide range of stencil sizes  which captures the behavior of such 3D stencil computation pattern using platform parameters ~\cite{DELACRUZ20112146}.

\newpage

\section{Minimod Description}
\label{sec:description}
\subsection{Source Code Structure}

In this section, we introduce the basic structure of the source code in Minimod. As we described in Section~\ref{sec:fd}, the simulation in Minimod consists of solving the wave equation, the temporal requires the spatial part of the equation to be solve at each timestep for some number of timesteps. The pseudo-code of the algorithm is shown in \autoref{algo:minimod}, for the second order isotropic constant density equation. We apply a Perfectly Matched Layer (PML) \cite{BERENGER1994185} boundary condition to the boundary regions. The resulting domain consists of an ``inner'' region where \autoref{eq:minimod-semidisc} is applied, and the outer ``boundary'' region where a PML calculation is applied.
\begin{algorithm}
	\KwData{$\mathbf{f}$: source}
	\KwResult{$\mathbf{p}^n$: wavefield at timestep $n$, for $n\leftarrow 1$ \KwTo $T$}
	$\mathbf{p}^0 := 0$\;
	\For{$n\leftarrow 1$ \KwTo $T$}{\nllabel{line:tsloop}
		\For{each point in wavefield $\mathbf{u}^n$}{
			Solve Eq.~\ref{eq:minimod-semidisc} (left hand side) for wavefield $\mathbf{p}^n$\;
		}
		$\mathbf{p}^n = \mathbf{p}^n + \mathbf{f}^n$ (Eq.~\ref{eq:minimod-semidisc} right hand side)\;
	}\nllabel{line:end-of-ts}
	\caption{Minimod high-level description}
	\label{algo:minimod}
\end{algorithm}

 As described in \autoref{algo:minimod}, the most computationally expensive component of minimod is the computation of the wavefield for each point. We list the code structure of the wavefield calculation in \autoref{algo:wavefield_solution}.
\begin{algorithm}
	\KwData{$p^{n-1}, p^{n-2}$: wavefields at previous two timsteps}
	\KwResult{$p^n$: wavefield at current timestep}
	\For{$i\leftarrow \mathrm{xmin}$ \KwTo $\mathrm{xmax}$}{\nllabel{line:xloop}
	    \eIf{$i\ge \mathrm{x3}$ {\bf and} $i \le \mathrm{x4}$}{
            \For{$j\leftarrow \mathrm{ymin}$ \KwTo $\mathrm{ymax}$}{
                \eIf{$j\ge \mathrm{y3}$ {\bf and} $j \le \mathrm{y4}$}{
                    \tcp{Bottom Damping (i, j, z1...z2)}
                    \tcp{Inner Computation (i, j, z3...z4)}
                    \tcp{Top Damping (i, j, z5...z6)}
                }{
                    \tcp{Back and Front Damping (i, j, zmin...zmax)}
                }
            }
	    }{
	        \tcp{Left and Right Damping (i, ymin...ymax, zmin...zmax)}
	    }
	}
	\caption{Wavefield solution step}
	\label{algo:wavefield_solution}
\end{algorithm}

\begin{figure}
    \centering
    \includegraphics[width=65mm]{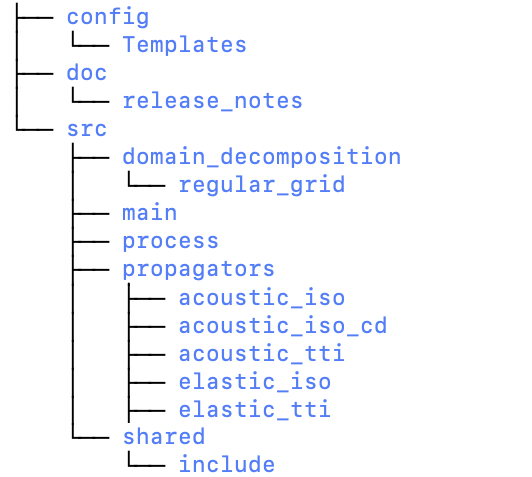}
\caption{Code tree structure of Minimod package.}
\label{fig:codetree}
\end{figure}

The general structure listed above is the backbone for all the propagators included in Minimod.
To keep the code simple and flexible, each propagator is compiled separately. This can be selected by setting the propagator variable in the version file before compiling. Figure~\ref{fig:codetree} presents a tree structure of Minimod code suite.

\subsection{Targets}

Each propagator has also its own implementation depending the hardware targeted. The target directory located in each propagator is the way of setting targets. In the source code structure, the following implementations of the target are provided:

\begin{itemize}

\item seq\_opt--none : implement kernels without any optimization (so as the short form of sequential none optimization mode). The purpose of the sequential implementation is to be used as a baseline to compare the optimized results. Also, this is useful to analyze not parallel-related characteristics.

\item omp\_opt--none : implement kernels using OpenMP directives for multi-threading execution on CPU. The goal of this target implementation is to explore the advanced CPU features on multi-core or many-core systems.

\item acc\_opt--gpu : implement kernels using OpenACC directives for offloading computing to GPU (so as the short form of accelerator optimization using GPU). The goal of this target implementation is to explore the accelerating on GPU using OpenACC programming model standard.

\item omp\_offload\_opt--gpu : implement kernels using OpenMP directives for offloading to GPU (so as the short form of OpenMP offloading optimization using GPU). The goal of implementing this target is to explore the accelerating on GPU using OpenMP programming standard.

\end{itemize}

In addition to change the propagator that is to be used in tests, one may also change the ``version" file to use a different target by setting the ``target" variable to the desired multi-threaded or accelerator implementations.

\subsection{Compilation and usage}

After the target propagators, compilers, and accelerating implementation settings are selected, the source code is ready for compilation, as follows:


     \begin{verbatim}
     To compile the sequential mode of Minimod package:
     $> source set_env.sh
     $> make 
     
     To compile the multi-threading mode with OpenMP directives:
     $> source set_env.sh
     $> make _USE_OMP=1
     
     To compile the offloading to GPU mode with OpenMP directives:
     $> source set_env.sh
     $> make _USE_OMP=tesla
     
     To compile the multi-threading mode with OpenACC directives:
     $> source set_env.sh
     $> make _USE_ACC=multicore
     
     To compile the offloading to GPU mode with OpenACC directives:
     $> source set_env.sh
     $> make _USE_ACC=tesla
     
     \end{verbatim}

The parameters of Minimod are shown in the following verbatim section. Those are the basic parameters for seismic modeling and they are set as command-line options. The main parameters include: \emph{grid sizes}, \emph{grid spacing} on each dimension, the number of \emph{time steps} and the maximum \emph{frequency}.

\begin{verbatim}
[]$ ./modeling_acoustic_iso_cd_seq_opt-none --help

 --ngrid                100,100,100   # Grid size
 --dgrid                20,20,20      # Dmesh: grid spacing
 --nsteps               1000          # Nb of time steps for modeling
 --fmax                 25            # Max Frequency
 --verbose              .false.       # Print informations
\end{verbatim}

In terms of expected results, the following verbatim section presents an example to show how to run the application and the run-time results of single-thread Minimod acoustic-iso-cd kernel. As we can see, the results report all the parameters that are used in the modeling and at the end the kernel time and modeling time of running the application.

\begin{verbatim}

[]]$ ./modeling_acoustic_iso_cd_seq_opt-none --ngrid 240,240,240 --nsteps 300

 nthreads           =            1
 
 ngrid              =          240         240         240
 dgrid              =    20.0000000       20.0000000       20.0000000    
 nsteps             =          300
 fmax               =    25.0000000    
 vmin               =    1500.00000    
 vmax               =    4500.00000    
 cfl                =   0.800000012    
 
 stencil            =            4           4           4
 source_loc         =          120         120         120
 ndamping           =           27          27          27
 ntaper             =            3           3           3
 
 nshots             =            1
 time_rec           =    0.00000000    
 nreceivers         =        57600
 receiver_increment =            1           1
 source_increment   =            1           1           0
 
 time step         100 /         300
 time step         200 /         300
 time step         300 /         300
Time Kernel       30.47
Time Modeling     31.01


\end{verbatim}

\section{Benchmarks}
\label{sec:benchmarks}
In this section examples of Minimod experimental results are presented. The purpose is illustrate performance and scalability of the propagators with regard to current HPC platforms.

\subsection{Experimental set-up}

The different propagators of Minimod are evaluated on Fujitsu A64FX architecture, AMD EYPC system, Intel Skylake and IBM Power8 system, as well as Nvidia's V100 GPUs. The specifications of hardware and software configurations of the experimental platforms are shown in Table~\ref{tab:hwsw}. 

\addtolength{\tabcolsep}{3pt}
\begin{table}
    \centering
    \begin{tabular}{l|c|c}
        \hline
        &\multicolumn{1}{|c|}{\textbf{Hardware}} & \multicolumn{1}{c}{\textbf{Software}}\\
        \hline
        CPUs & A64FX Armv8-A SVE architecture & Fujitsu Compiler 1.1.13 (frt)\\
        CPU cores & 48 computing cores & OpenMP (-Kopenmp) \\
        Memory & 32 GB HBM2 & auto-parallelisation\\
        L2 & 8 MB & (–Kparallel)\\
        L1 & 64 KB & \\
        Device Fabrication & 7nm & \\
        TDP & 160W & \\ 
        \hline

        CPUs & AMD EYPC 7702 & GCC 8.2 (gfortran)\\
        CPU cores & 64 computing cores & OpenMP \\
        Memory & 256 GB & \\
        L3 & 256 MB (per socket) & \\
        L2 & 32 MB & \\
        L1 & 2+2 MB & \\
        Device Fabrication & 14nm & \\ 
        TDP & 200W & \\ 
        \hline
        
        CPUs & 2x Intel Xeon Gold 5118 & intel compiler 17.0.2 (ifort)\\
        CPU cores & 24 (12 per CPU)  \\
        Memory & 768 GB & \\
        L3 & 16 MB (per socket) & \\
        L2 & 1024 KB & \\
        L1 & 32+32 KB & \\
        Device Fabrication & 14nm & \\ 
        TDP & 2 x 105W & \\ 
        \hline

        CPUs & 2 x IBM Power8 (ppc64le) &  PGI 19.7 (pgfortran) \\
        CPU cores & 20 computing cores (10 per CPU) & OpenMP (-mp) \\
        Memory & 256 GB & \\
        L3 & 8 MB (per two cores) & \\
        L2 & 512 KB (per two cores) & \\
        L1 & 64+32 KB & \\
        Device Fabrication & 22nm &\\
        TDP & 2 x 190W & \\
        \hline
        
    
        GPU & 1 x Nvidia V100 &  PGI 19.7 (pgfortran) \\
        cores & 2560 Nvidia CUDA cores & OpenACC (-ta=tesla) \\
        Memory & 16 GB HBM2& \\
        L2 & 6144 KB & \\
        Device fabrication & 12nm FFN &\\
        Power consumption & 290W & \\ 
        \hline
        
    \end{tabular}
    \caption{Hardware and software configuration of the experimental platforms. From top to bottom, the first section is Fujitsu A64FX Arm8-A architecture. The second section is AMD EYPC Rome architecture. The third section is Intel Skylake architecture. The fourth section is IBM Power8 architecture. And the bottom section is the specification of Nvidia's V100 GPU.}
    \label{tab:hwsw}
\end{table}

\newpage

\subsection{Performance characterization}

In our experiments, we use roofline model proposed by Williams et al. \cite{roofline2009} to understand the hardware limitations as well as evaluating kernel optimization. In the roofline model, the performance of various numerical methods are upper bounded by the peak floating point operations (flop) rate and the memory bandwidth while running on single-core, multi-core or accelerator processor architecture. 

\begin{figure}[ht!]
    \centering
    \includegraphics[width=150mm, height=80mm, scale=0.5]{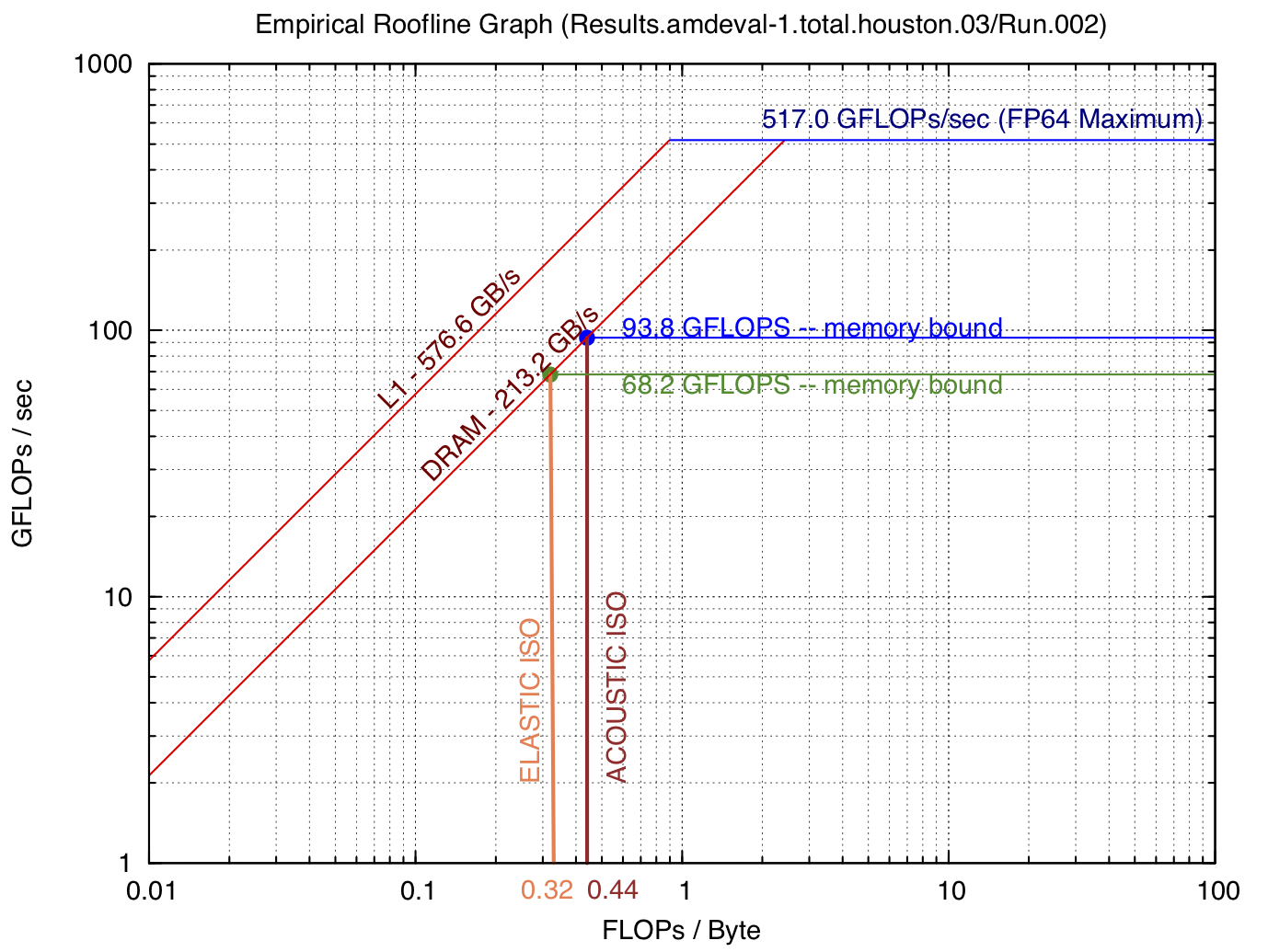}
\caption{Roofline model analyses on AMD EYPC System.}
\label{fig:roofline}
\end{figure}

Figure~\ref{fig:roofline} shows the peak performance in term of GFlops per seconds and memory bandwidth of cache and main dynamic random-access memory (DRAM) of the AMD EYPC system listed in Table~\ref{tab:hwsw} where we conducted experiments on. The arithmetic intensity in the roofline plot is calculated by the number of floating point operations that are performed in the stencil calculation divided by the number of words that we need to read from and write to memory~\cite{delacruz2014semi}.

\subsection{Single compute node-level parallelism}

\begin{figure}[ht!]
    \centering
    \includegraphics[scale=0.65]{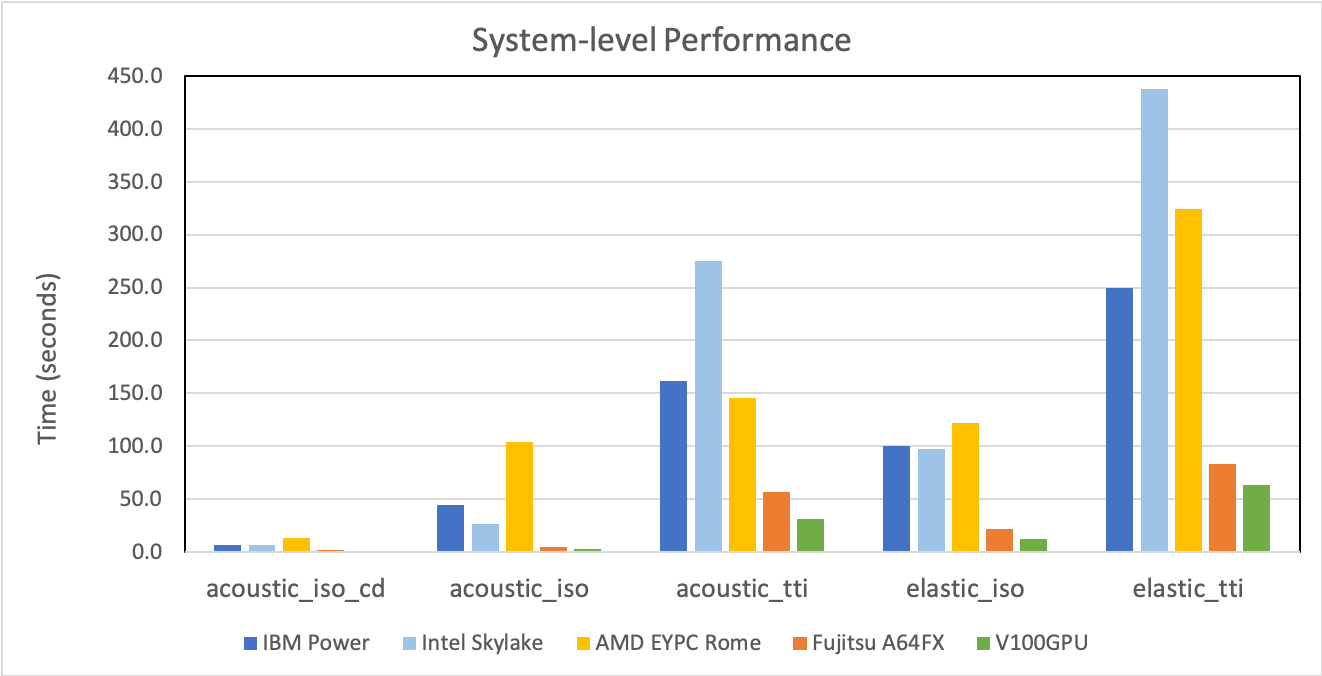}
\caption{System-level performance comparison results of Minimod with different propagators running on IBM power system (dark blue bars), on Intel Skylake system (light blue bars), on AMD EPYC system (yellow bars), on Fujitsu A64FX system (orange bars), and on NVIDIA's V100 GPU (green bars).}
\label{fig:systemperf}
\end{figure}

\begin{figure}[tb!]
    \centering
    \includegraphics[scale=0.5]{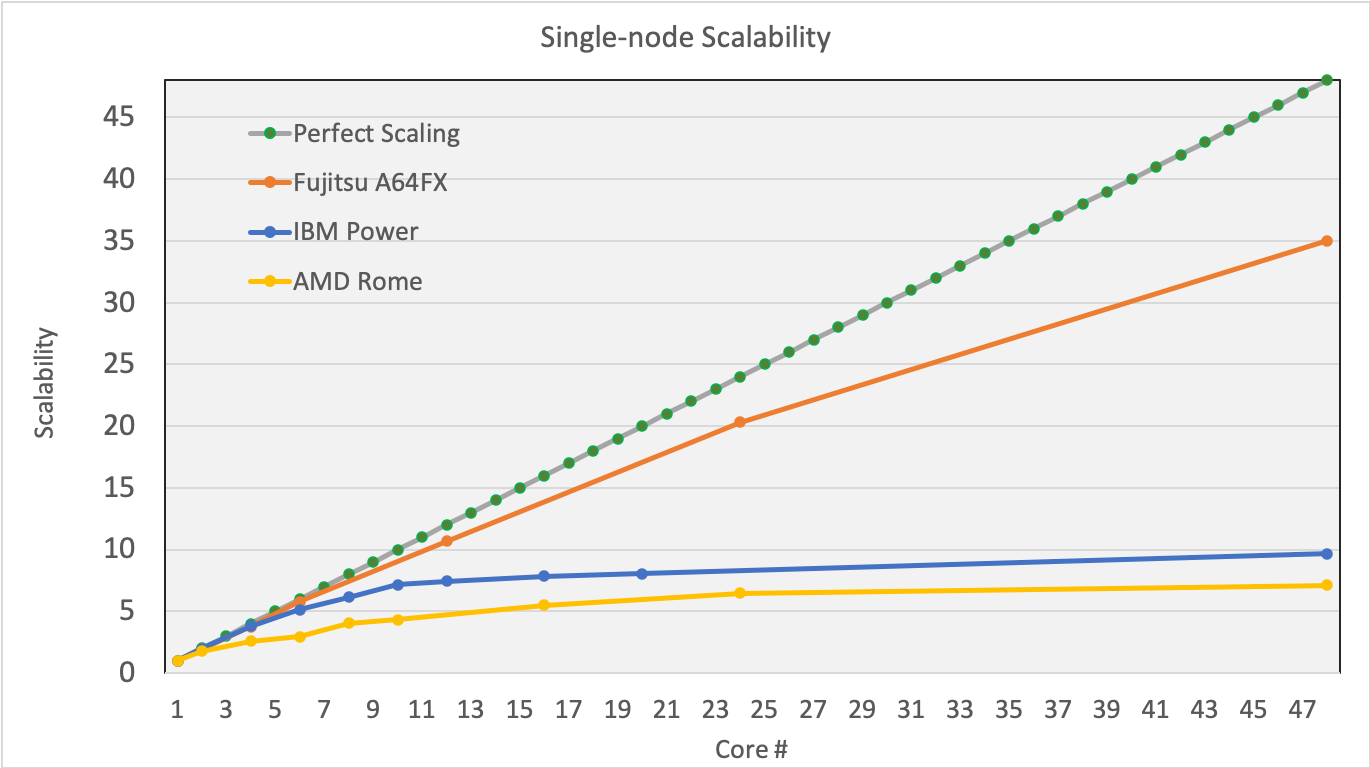}
\caption{Single compute node-level scalability comparison results of Minimod running on IBM power system (blue curve), on AMD EYPC Rome system (yellow curve), and on Fujitsu A64FX system (red curve), and both are compared against the ideal scale-up (green curve).}
\label{fig:nodestrscal}
\end{figure}

We use Minimod to experiment the single compute node-level parallelism on different computer systems. As shown in Figure~\ref{fig:nodestrscal}. The system-level performance tests are conducted on IBM power, Fujitsu A64FX systems, and compared with using NVIDIA's V100 GPUs as accelerators. The different propagators in Minimod (acoustic\_iso\_cd, acoustic\_iso, acoustic\_tti, elastic\_iso, and elastic\_tti) are tested, and results are shown in Figure~\ref{fig:systemperf}. \\

As we observe in Figure~\ref{fig:systemperf}, the Fujitsu A64FX processor (as shown in the orange bars) provides better performance for all the propagators in comparison to both IBM power system (as shown in the dark blue bars), Intel skylake system (as shown in the light blue bars), as well as AMD EYPC Rome systems (as shown in the yellow bars). In fact, the performance of Fujitsu A64FX is closer to the performance of the system with Nvidia's V100 GPU accelerator (as shown in the green bars). \\

The single node-level scalability tests are conducted on IBM power, AMD EYPC, and Fujitsu A64FX systems. The problem size for the strong scalability tests are set to be 240 x 240 x 240. As presented in Figure~\ref{fig:nodestrscal}, the results are compared between the three modern computer systems and also compares against the ideal case. Across the three systems, Fujitsu A64FX system again wins IBM power and AMD EYPC Rome systems in the single-node scalability tests. 

\subsection{Distributed Memory Approach}

\begin{figure}[ht!]
    \centering
    \includegraphics[width=\textwidth]{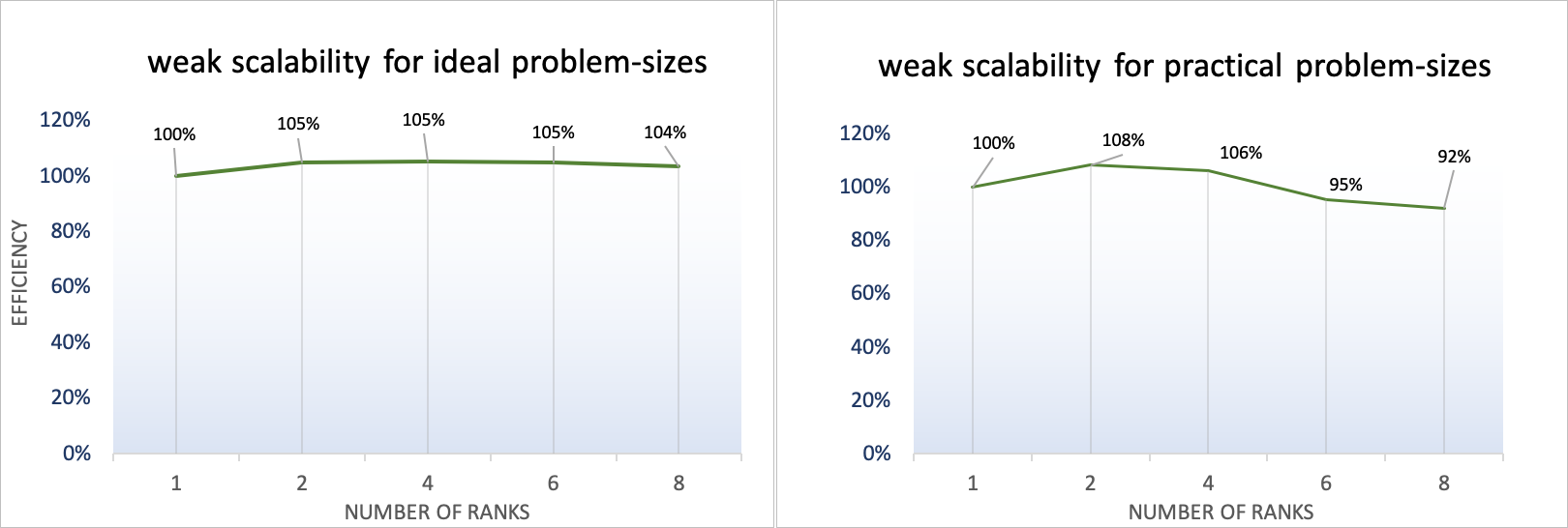}
\caption{MPI weak scalability of Minimod running on IBM power system for both the ideal problem-sizes (baseline $1000 \times 1000 \times 1000$ with linear increment on X dimension) and practical problem-sizes (the problem-size   increments are balanced on each dimension) running on 1 to 8 MPI ranks respectively.}
\label{fig:mpiweakscal}
\end{figure}

The distributed version of Minimod is implemented using Message Passing Interface (MPI). The domain decomposition is defined using regular Cartesian topology, and the domain decomposition parameters need to match the total number of MPI ranks: for example, for the three-dimensional domain decomposition in $x\times y\times z$ equals $2\times 2\times 4$, the rank number needs to be $16$. As for the time being, only \emph{acoustic\_iso\_cd} propagator is available within the distributed version of Minimod. \\

\begin{figure}[hb!]
    \centering
    \includegraphics[scale=0.5]{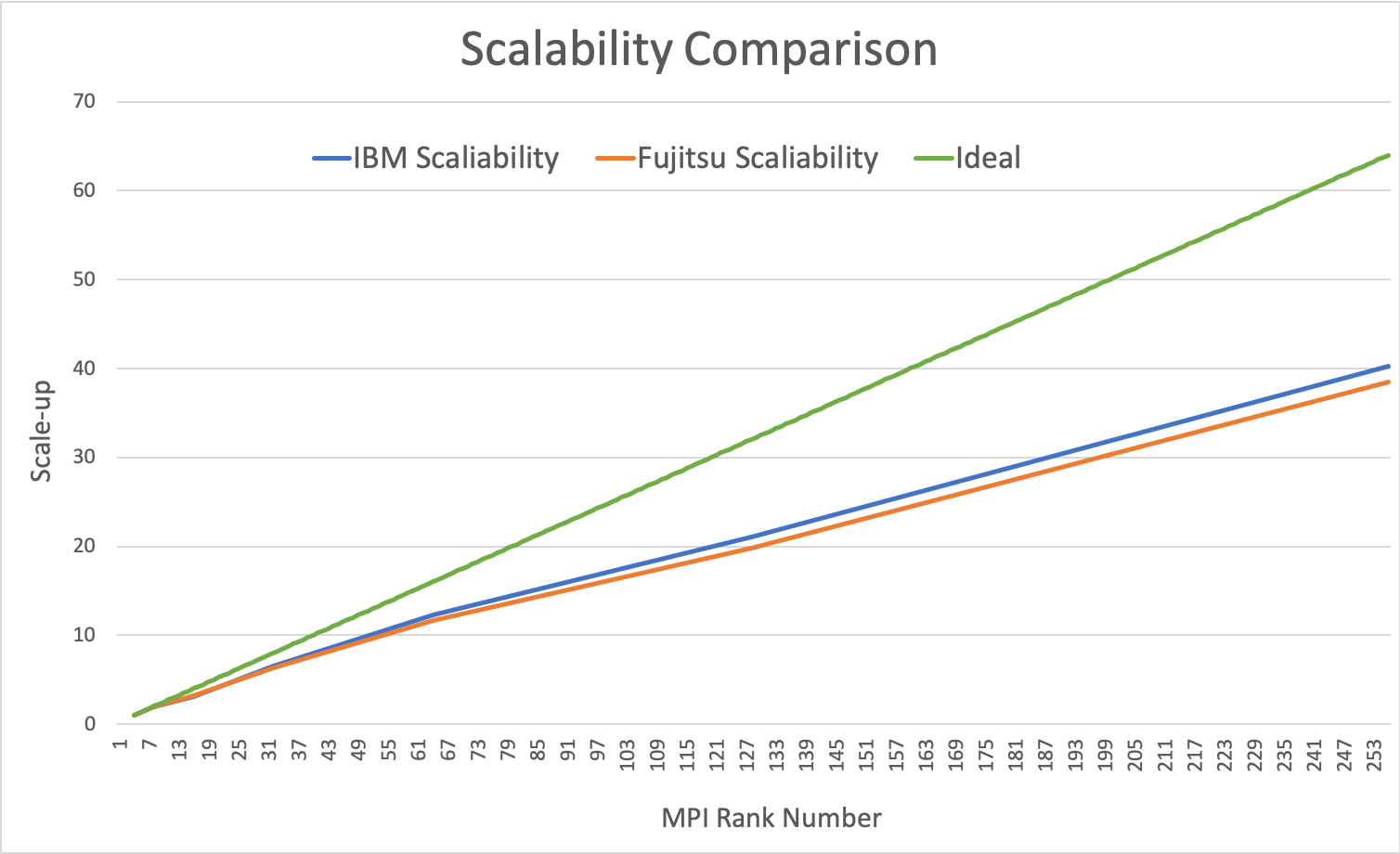}
\caption{MPI strong scalability comparison results of Minimod running on IBM power system (blue curve) and on Fujitsu A64FX system (red curve), and both are compared against the ideal scale-up (green curve).}
\label{fig:mpistrscal}
\end{figure}

The implementation of MPI communication between sub-domains uses non-blocking send and receives. The communication operates in ``expected message" mode that has no overlap of communication with computation. Each subdomain performs the following steps: first, to pack the messages to be transmitted in buffers; second, to perform communication by posting all sends and receives, and finally wait till the communication is complete and unpacks the arrived data.\\

We evaluated both weak scalability and strong scalability of the distributed version of Minimod for \emph{acoustic\_iso\_cd} propagator. The results of weak scalability running Minimod on IBM power system is shown in Figure~\ref{fig:mpiweakscal}, which presents the evaluation results using both the ideal problem sizes and the practical problem sizes running on 1 to 8 MPI ranks, respectively.\\

For the weak scalability test running ideal problem sizes, we used a baseline of $1000 \times 1000 \times 1000$ with linear increment on X dimension (for example, for the test running on 6 MPI ranks we used a problem-size of $6000 \times 1000 \times 1000$). And for practical problem-sizes, we used the same baseline while the problem-size increments are balanced on each dimension (for example, for the test running on 6 MPI ranks we used a problem-size of $1856 \times 1856 \times 1856$). The green curves in Figure~\ref{fig:mpiweakscal} present the efficiencies in comparison to the baseline result. \\

We can see from Figure~\ref{fig:mpiweakscal} that the weak scalability holds well for running from 1 rank scale to up to 8 ranks for the ideal problem sizes. And for the practical problem sizes which is more close to the real seismic acquisition situation, the weak scalability efficiencies for 2 ranks and 4 ranks are higher than 100\% because of the slightly smaller problem sizes compared to the baseline case ($1280 \times 1280 \times 1280$ for 2 ranks and $1600 \times 1600 \times 1600$ for 4 ranks), while it starts diminishing when it reaches 8 ranks mainly because of the increase of problem sizes.\\

The results of strong scalability are shown in Figure~\ref{fig:mpistrscal}. The strong scalability tests are conducted on both IBM power and Fujitsu A64FX systems. The problem size for the strong scalability tests is set to $1024 \times 1024 \times 1024$, on the rank numbers of 8, 16, 32, 64, 128, and 256 respectively. \\

As presented in Figure~\ref{fig:mpistrscal}, the results of the kernel execution on the IBM power and the Fujitsu A64FX systems are compared with the ideal scaling trend. The strong scalability results on both systems are very close when the MPI rank number is smaller than 64, while the kernel shows slightly better scalability results on the IBM system than on the Fujitsu system when running with 128 and 256 MPI ranks. In comparison to the ideal case, scalability on the IBM power system reached 63\% while on the Fujitsu system reached 60\% of the ideal scalability. \\

\subsection{Profiling}
\begin{figure}[ht!]
    \includegraphics[width=\textwidth, height=4.5in]{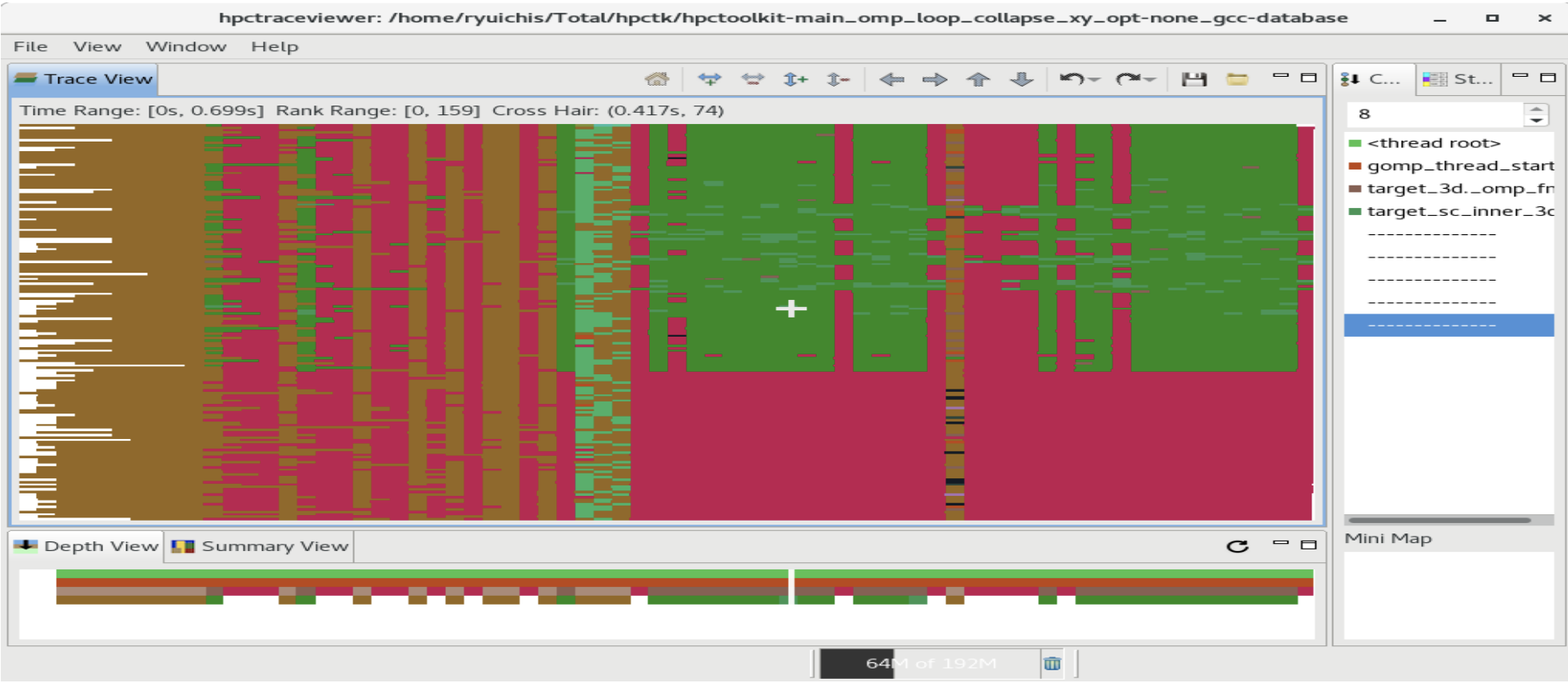}
\caption{Profiling results of Minimod using HPCToolkit.}
\label{fig:profiling}
\end{figure}

Profiling and analyses was conducted on Minimod, for example using the HPCToolkit~\cite{MC2001} from Rice University. Figure~\ref{fig:profiling} shows a screenshot of the trace view in HPCToolkit profiling Minimod acoustic iso kernel implemented in multi-threading mode using OpenMP. The biggest panel on the top left   presents sequences of samples of each trace line rendered. The different colors represent the time spends on different subroutines which are listed on the right panel. The bottom panel in Figure~\ref{fig:profiling} is the depth view of the target Minimod application which presents the call path at every time step.  \\

\begin{figure}[ht!]
    \hspace{-.23in}
    \includegraphics[width=6.9in, height=5in]{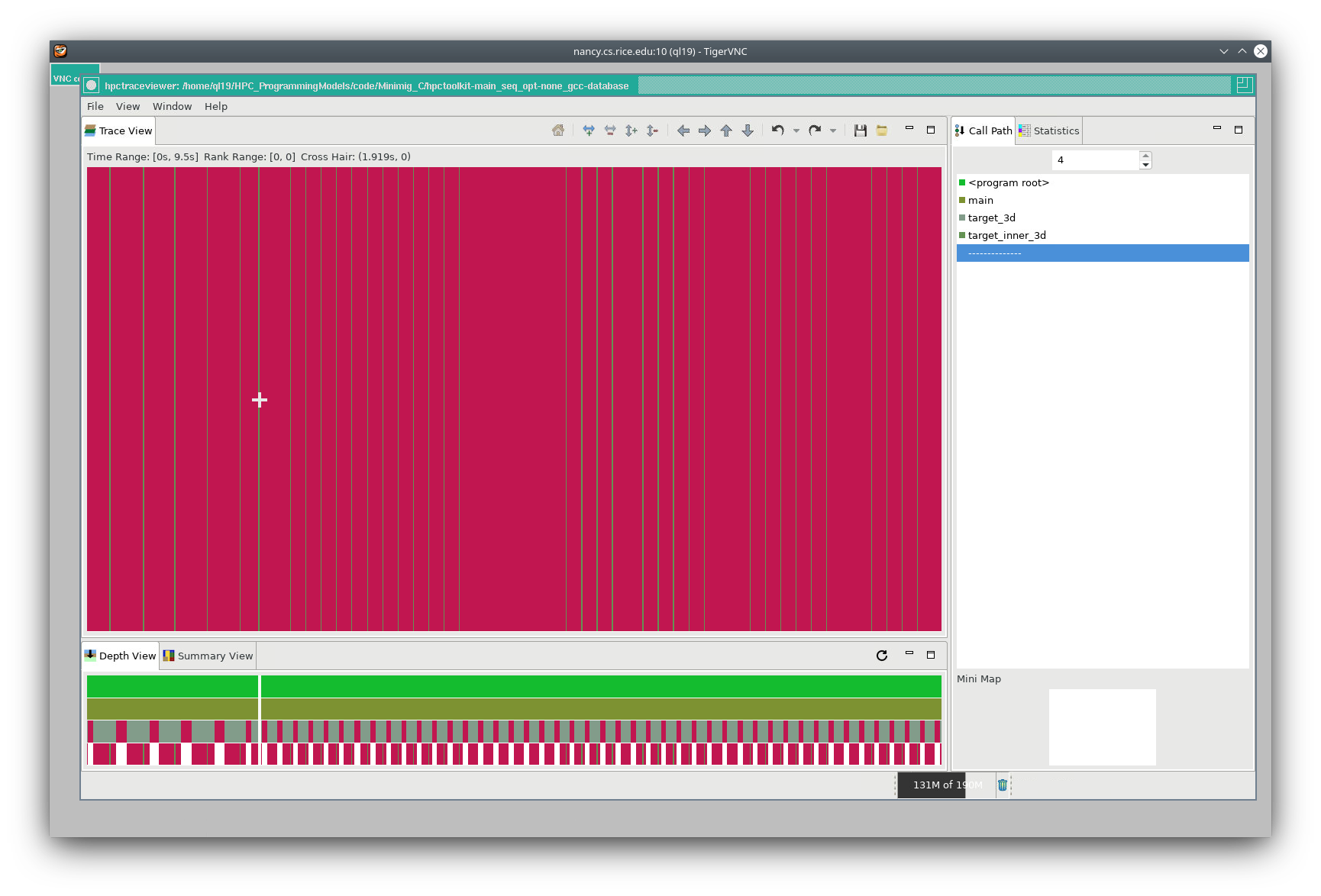}
\caption{Example of profiling sequential mode of Minimod acoustic-iso-cd kernel using HPCToolkit.}
\label{fig:profilingzoom}
\end{figure}

As an illustrative example for profiling Minimod, Figure~\ref{fig:profilingzoom} shows the profiling results from HPCToolkit trace view for the sequential implementation of the simplest kernel acoustic\_iso\_cd (acoustic wave-propagation in isotropic media with constant density) in Minimod without any optimization. To better understand the behavior of the kernel, what is shown in the picture is a case with one thread with the stencil computation on a grid size of $100\times100\times100$. As shown in Figure~\ref{fig:profilingzoom}, the majority of the time is spent on running the ``target\_pml\_3d" which is the implementation of perfectly-matched region, as shown in the dark color areas in the top left panel. And the green vertical line is for the ``target\_inner\_3d", where the thread performs computation for the inner region of stencil. \\

\begin{figure}[ht!]
    \centering
    \includegraphics[width=\textwidth, height=3.5in]{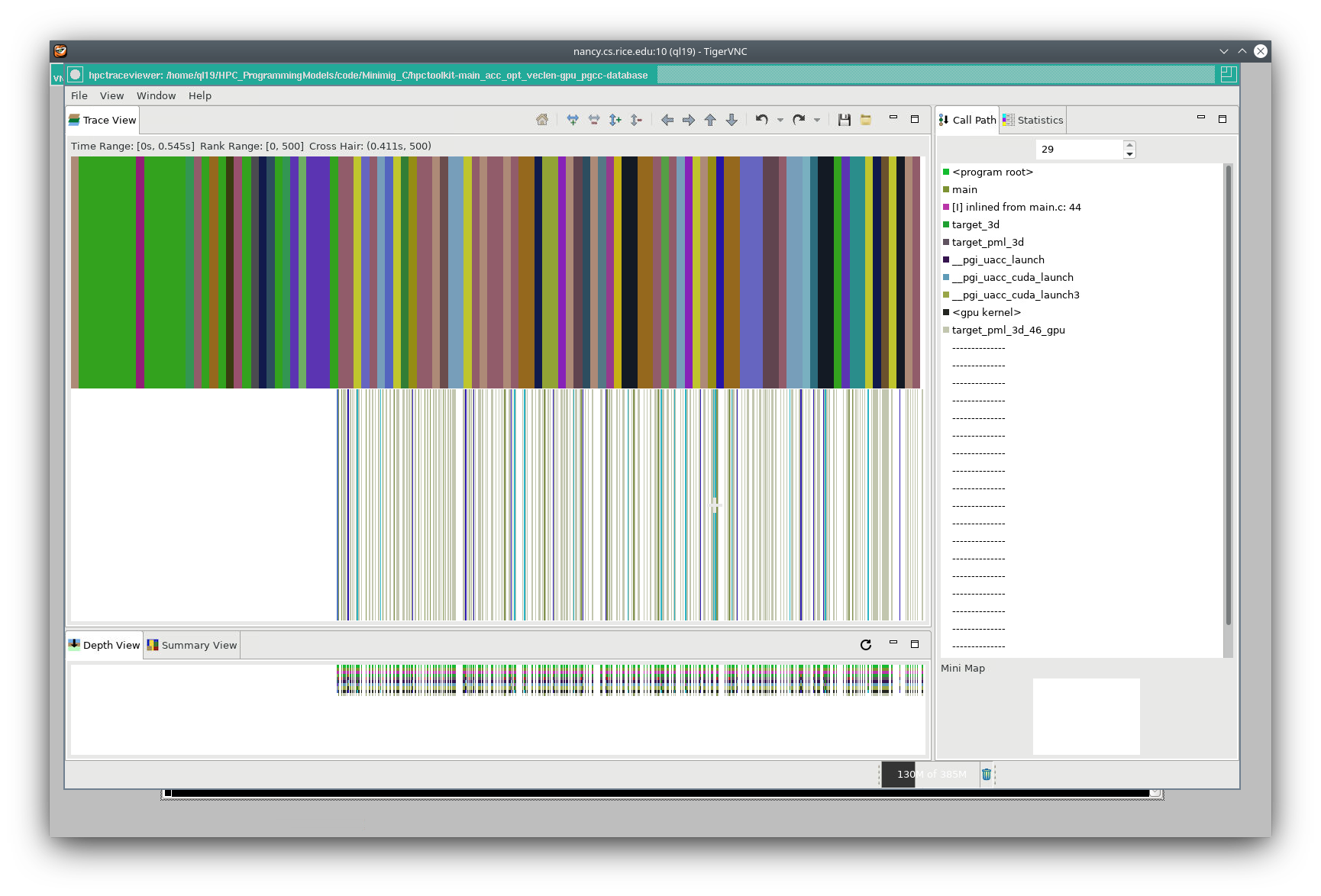}
\caption{GPU profiling results of Minimod acoustic-iso-cd kernel using HPCToolkit.}
\label{fig:gpuprofiling}
\end{figure}

An advantage of HPCToolkit it that can profile the results of Minimod GPU mode for each time sampling traces. Figure~\ref{fig:gpuprofiling} shows the the profiling results of the entire execution of Minimod acoustic-iso-cd kernel in OpenACC offloading to GPU mode. Different than the CPU profiling trace views, the GPU profiling trace view on HPCToolkit top-left panel window is composed of two rows. The top row shows the CPU (host) thread traces and the bottom row is for the GPU (device) traces. \\

\begin{figure}[bt!]
    \centering
    \includegraphics[width=\textwidth, height=3.5in]{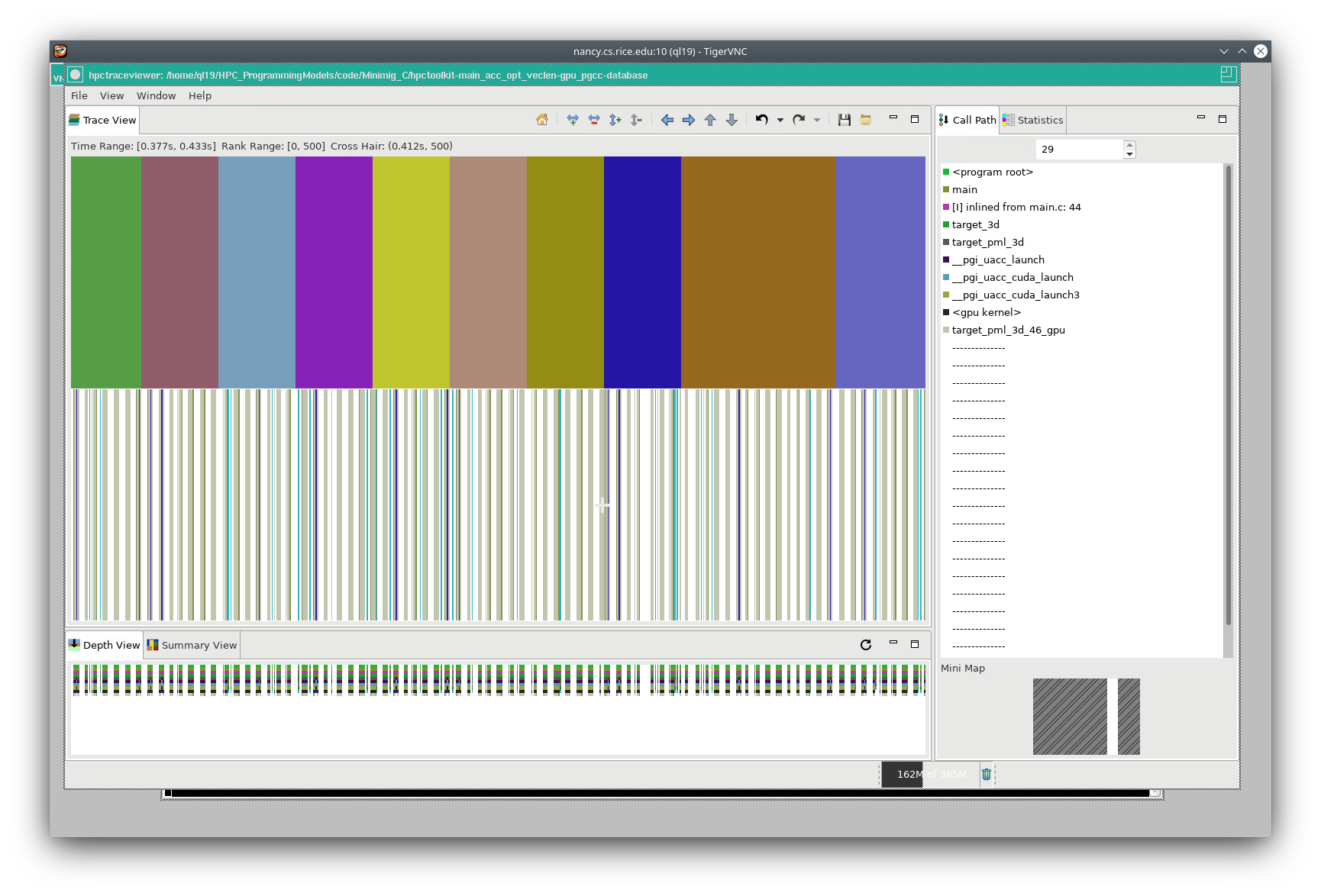}
\caption{A detailed view of GPU profiling Minimod acoustic-iso-cd kernel using HPCToolkit.}
\label{fig:gpuprofilingzoomin}
\end{figure}

A zoomed-in view of this GPU profiling results is presented in Figure~\ref{fig:gpuprofilingzoomin}. We selected time step shows the GPU that is running the ``target\_pml\_3d" kernel where the blank white spaces in the GPU row shows the idleness. The same as in the profiling results for CPU, different colors here represent the time spends on different GPU calls.

\newpage

\section{Conclusion}
\label{sec:conclusion}
This article introduces a proxy application suite for seismic modeling using finite difference method named \emph{Minimod}. The design concepts, underline algorithms, and code structures of Minimod are described. The benchmark results of Minimod are shown on different computer architectures for both single compute node-level parallelism  and distributed memory approaches. 

\section{Acknowledgements}
We would like to thank Total and subsidiaries for allowing us to share this material. We would also like to express our appreciation to Diego Klahr for his continuous support, and our colleague Elies Bergounioux in France for discussions on the adaptability of proxy applications in production. We also thank Ryuichi Sai from Rice University for his contribution on the profiling results using HPCToolkits. We would like acknowledge Pierre Lagier from Fujitsu for his help with the experiments conducted with latest Fujitsu technology. Last but not least, many thanks to our former colleague Maxime Hugues for his initial implementation of the presented software.

\bibliographystyle{abbrv}
\bibliography{ref}

\end{document}